\documentclass[aps,pre,showpacs,twocolumn,preprintnumbers,amsmath,amssymb,showkeys]{revtex4-1}

\usepackage{graphicx}
\usepackage{dcolumn}
\usepackage{bm}

\newcommand\bk{\textbf{k}}

\begin{document}



\title{A systematic bias in the calculation of spectral density from a 3D spatial grid}


\author{Rodion Stepanov}
\email[]{rodion@icmm.ru}
\affiliation{Institute of Continuous Media Mechanics, Korolyov str.\ 1,
614013 Perm, Russia}
\affiliation{Perm National Research Polytechnic University, Komsomolskii av. 29, 614990 Perm, Russia}
\author{Franck Plunian}
\affiliation{Univ. Grenoble Alpes, ISTerre, F-38000 Grenoble, France}
\affiliation{CNRS, ISTerre, F-38000 Grenoble, France}
\author{Mouloud Kessar and Guillaume Balarac}
\affiliation{Univ. Grenoble Alpes, LEGI, F-38000 Grenoble, France}
\affiliation{CNRS, LEGI, F-38000 Grenoble, France}

\date{\today}

\begin{abstract}
The energy spectral density $E(k)$, where $k$ is the spatial wave number, is a well-known diagnostic of homogeneous turbulence and magnetohydrodynamic turbulence.
However in most of the curves plotted by different authors, some systematic kinks can be observed at $k=9$, $k=15$ and $k=19$. We claim that these kinks have no physical meaning, and are in fact the signature of the method which is used to estimate $E(k)$ from a 3D spatial grid.
In this paper we give another method, in order to get rid of the spurious kinks and to estimate $E(k)$ much more accurately.
\end{abstract}

\maketitle

\section{Motivation}
Assuming isotropic and homogeneous hydrodynamic turbulence, Kolmogorov predicted that kinetic energy spectral density should have an universal power scaling of $k^{-5/3}$ \cite{Kolmogorov1941}, where $k$ is the spatial wave number. Since then, the energy spectral density became a useful diagnostic tool for various configurations, including anisotropic and magnetohydrodynamic turbulence.
The definition of the spectral density $E(k)$ is given by \citep{Lesieur1997}
\begin{equation}
E(k)=\int_{|{\bf k'}|=k}\hat{E}({\bf k'})\, d{\bf k'}, \label{eq1}
\end{equation}
where $\hat{E}({\bf k'})$ is the Fourier transform of the autocorrelation of a scalar field or the trace of the autocorrelation tensor of a vector field, e.g. velocity or magnetic field \citep{Tennekes}.

\begin{figure*}
\begin{center}
\includegraphics[width=0.8\textwidth]{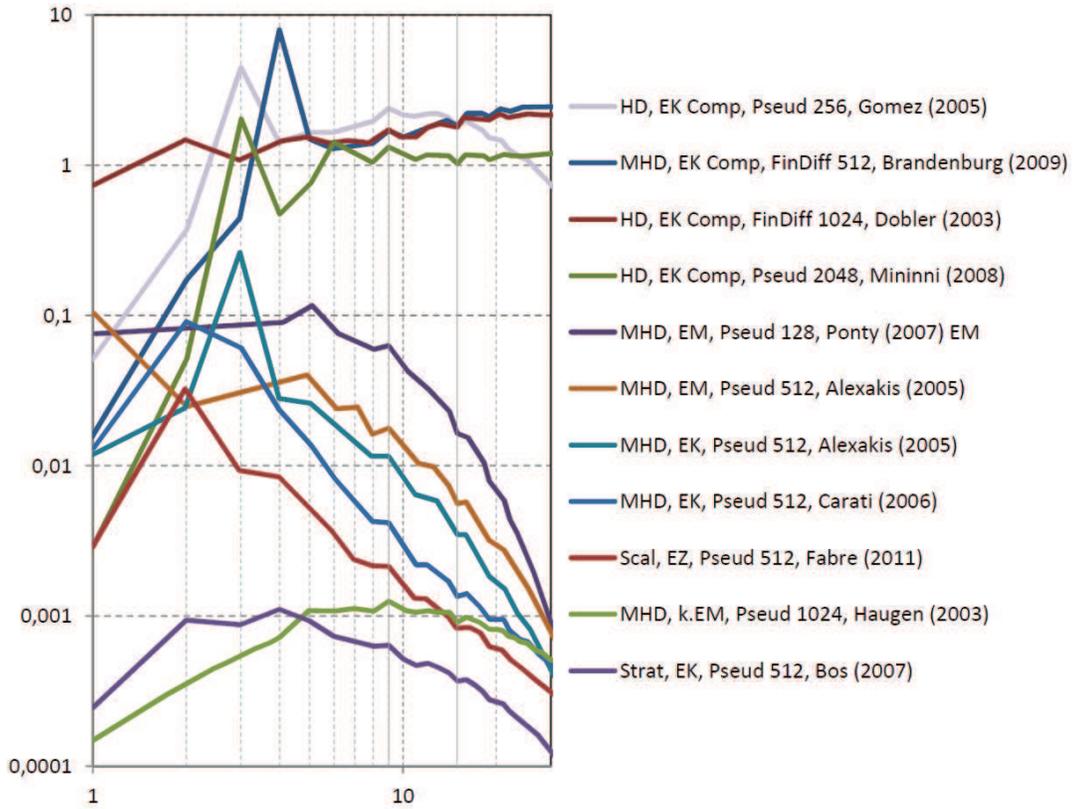}
\end{center}
\caption{Energy spectra calculated by different authors, with different codes, different resolutions and for different quantities. The notations HD, MHD and Scal stand for hydrodynamic, magnetohydrodynamic and passive-scalar problems. The notations EK, EM and EZ stand for kinetic, magnetic and passive-scalar energies. Pseud and FinDiff stand for pseudo-spectral and finite-difference resolution methods, the numbers 128, 512, 1024 and 2048 correspond to the spatial resolution. The vertical grid lines $k=9$ and $k=15$ are highlighted as solid lines.}
\label{fig:turbspectra}
\end{figure*}
We look at  definition (\ref{eq1}) from the point of view of turbulence in a computational box. In practice the numerical implementation of (\ref{eq1}) for $\hat{E}({\bf k'})$ given on a regular grid of mesh is not discussed. However some common features can be distinguished in the results.
As an example, a compilation of curves corresponding to kinetic, magnetic and passive-scalar energy spectra obtained in hydrodynamic or magnetohydrodynamic turbulence is plotted in Figure \ref{fig:turbspectra}.
Though these spectra have been obtained by various authors \citep{Alexakis2005,Bos2007,Brandenburg2009,Carati2006,Dobler2003,Fabre2011,Gomez2005,Haugen2003,Mininni2008,Ponty2007}, using various methods, forcing and degrees of resolution, we note a systematic  bump at scale $k=9$, followed by two holes at scale $k=15$ and $k=19$. Bumpy spectra are familiar in the context of wave turbulence, usually interpreted as the signature of travelling modes \cite{Figueroa2013}, but they are found in time frequency only. Here however it is difficult to imagine any physical ground for the systematic $k=9$, $k=15$ and $k=19$ kinks appearing in the inertial range. It is fair to say that usually these kinks are just ignored in discussions of physical or even numerical aspects of the results \citep{2003JCoPh.184..366C,Desjardins20087125}, even if the bump at $k=9$ has already been interpreted as a physical effect \citep{Haugen2003}. Another hole at $k=3$ might be found but it is usually hidden by the forcing scales.

In section \ref{sec:from} we show that, in fact, these kinks are produced by a systematic bias coming from the
standard approach to estimate $E(k)$ on a spatial grid. In section \ref{sec:bias} we present a new method to estimate $E(k)$ in order to circumvent this bias.
Such a bias being more striking in 3D than in 2D turbulence, in this paper we consider  only the case of 3D data sets. A new definition for the 2D case is however given in section \ref{Conclusion}. In 1D models like EDQNM models \citep{Pouquet1976} or shell models \citep{Plunian2013} of turbulence, this bias does not exist.

\section{Where does the bias come from?}
\label{sec:from}

\begin{figure}
\begin{center}
\includegraphics[width=0.49\textwidth]{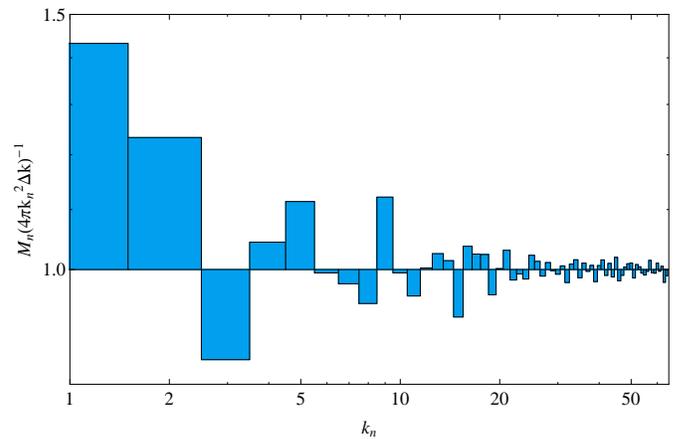}
\end{center}
\caption{Density of the number of vectors belonging to shell $S_n$ versus $k_n$, for $\Delta k=\delta k$.}
\label{fig:Number per shell}
\end{figure}

The standard approach to estimate the continuous quantity $E(k)$ from a set of Fourier modes given on a regular grid of mesh $\delta k$, is to divide the Fourier space in shells $S_n$ of thickness $\Delta k$.
Then the spectral density $E_n$ can be defined as \citep{Lesieur1997}
\begin{equation}
E_n = \frac{(\delta k)^3}{\Delta k}\sum_{\bk'\in S_n} \hat{E}({\bk'}), \label{Ekn}
\end{equation}
with
\begin{equation}
S_n=\{\bk' \in \mathbb{R}^3 \; / \; n\Delta k-\Delta k/2 < |\bk'| \le n\Delta k+\Delta k/2\}.\label{Sn}
\end{equation}
Usually it is natural to take $\Delta k=\delta k=1$, leading to a
unity pre-factor in (\ref{Ekn}). The wave number $k_n$
corresponding to shell $S_n$ is usually taken to obey an
arithmetic progression. Then it is defined as
\begin{equation}
k_n=n\Delta k.\label{kn}
\end{equation}
The problem is that the number $M_n$ of wave vectors $\bk'$ belonging to $S_n$ is not exactly proportional to the shell volume, as depicted in Figure \ref{fig:Number per shell} for $\Delta k=\delta k$. The density of $M_n$ even reaches local extrema at $k_n=9$, $k_n=15$ and $k_n=19$, which clearly explains the kinks appearing in Figure \ref{fig:turbspectra}.
Changing the value of $\Delta k$ would not help.
For $\Delta k < \delta k$ the number of local extrema is getting larger, while taking
$\Delta k > \delta k$ leads to a spurious power law as the result of an over averaging procedure.
In Figure~\ref{fig:Number per shell} we note another bump at $k=5$ which presumably is also responsible for the peaks at $k=5$ visible in several spectra of Figure~\ref{fig:turbspectra}, like
the one calculated by  \citet{Ponty2007}.

\section{How to circumvent the bias?}
\label{sec:bias}
Starting from (\ref{eq1}), we note that $E(k)$ is a surface integral, over $|{\bf k'}|=k$.
Keeping in mind that the surface of a shell of radius $|\bk'|$ is equal to $4\pi |\bk'|^2$,
we then introduce the following definition for the spectral density in shell $S_n$, now denoted $E_n^{*}$, in the form
\begin{equation}
E_n^{*}=\frac{4\pi}{M_n} \sum_{\bk'\in S_n}  \hat{E}(\bk')|\bk'|^2, \label{Eknstar}
\end{equation}
where again $M_n$ is the number of vectors $\bk'$ belonging to shell $S_n$.

In the same line, in order to estimate the mean wave number in shell $S_n$, we suggest to simply average all wave numbers belonging to $S_n$. This average wave number, now denoted $k_n^*$, is given by
\begin{equation}
k_n^*=\frac{1}{M_n}\sum_{\bk' \in S_n}|\bk'|. \label{knstar}
\end{equation}

Finally as we are looking for an energy spectral density
satisfying some power law it makes sense to use a geometric
progression, instead of an arithmetic one, for $k_n$. Then the
shells are logarithmically spaced ($n\propto\log k_n$), instead of
being linearly spaced ($n\propto k_n$). Then we define the new
shells as,
\begin{equation}
S_n^{log}=\{\bk' \in \mathbb{R}^3 \; / \; \lambda^n\delta k <
|\bk'| \le \lambda^{n+1} \delta k\},\label{Snbis}
\end{equation}
where $\lambda$ is some scalar value larger than unity.

In order to test our new definitions $k_n^*$ and $E_n^{*}$, we consider a synthetic set of data with spectral coefficients $\hat{E}(\bk')=\frac{1}{4\pi}|\bk'|^{-11/3}$. It corresponds to the exact spectral density $E(k)=k^{-5/3}$.

\begin{figure}
\begin{center}
\includegraphics[width=0.49\textwidth]{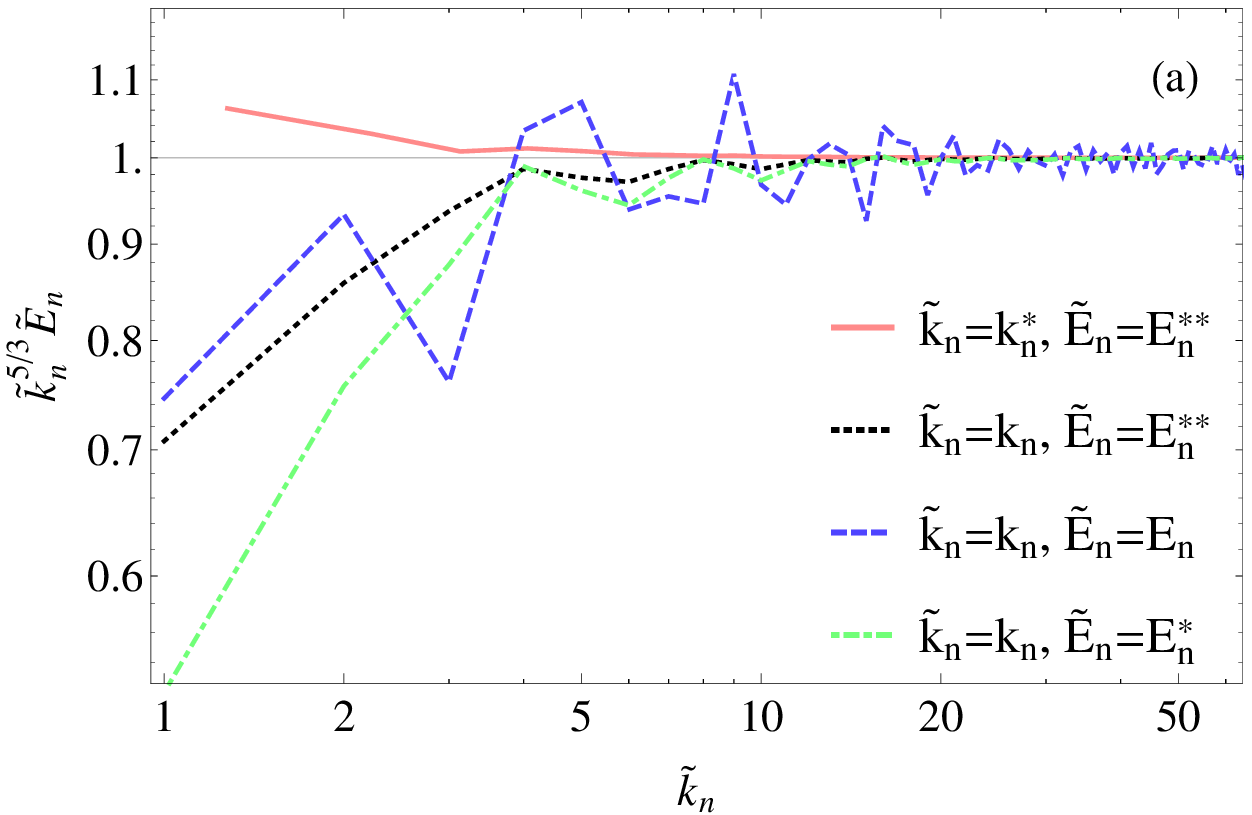}
\includegraphics[width=0.49\textwidth]{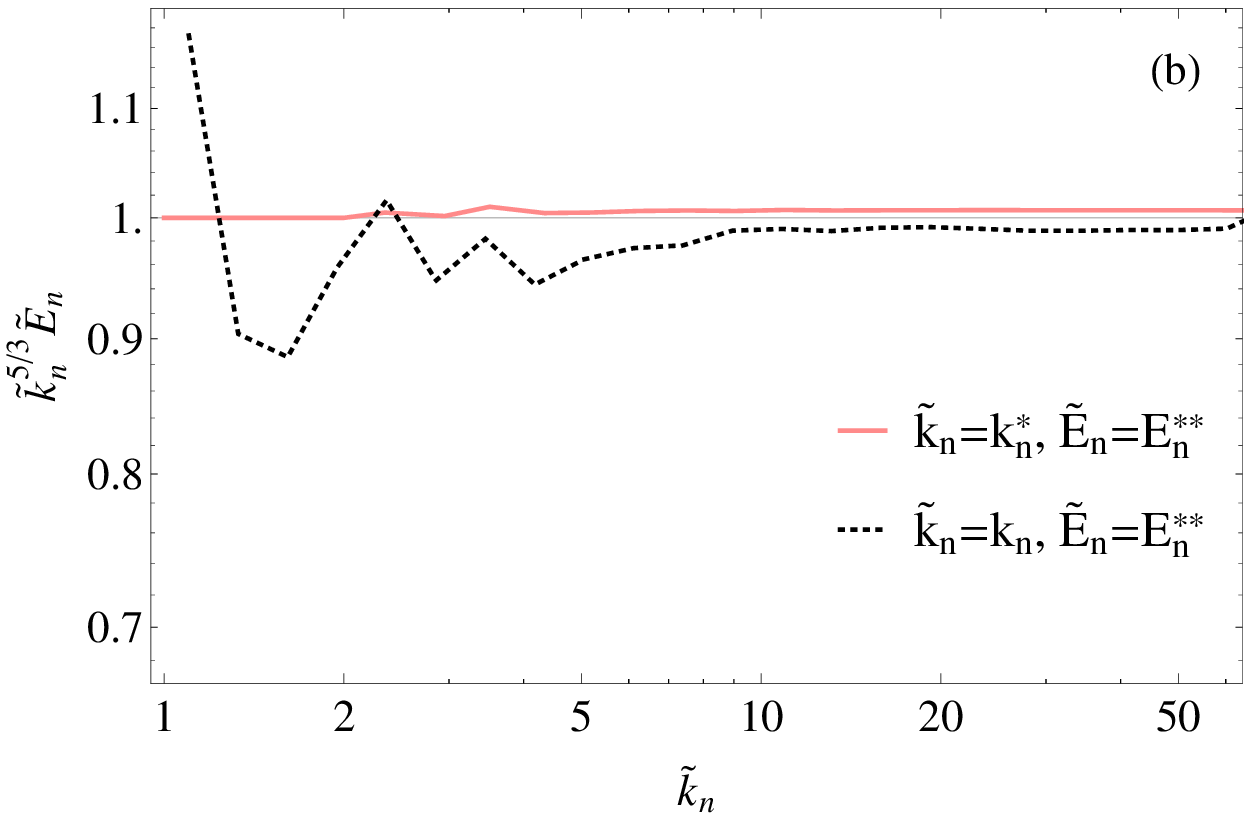}
\end{center}
\caption{Compensated spectral density
$\tilde{k}_n^{5/3}\tilde{E}_n$ versus $\tilde{k}_n$, (panel a) for
linearly spaced shells, $S_n$, and (panel b) logarithmically
spaced shells, $S_n^{log}$. In the panel (a), the four curves
correspond to $\tilde{k}_n=k_n$ or $k_n^*$, and $\tilde{E}_n=E_n$
or $E_n^{*}$. In the panel (b), the curves corresponds to
$\tilde{E}_n=E_n^{*}$ and $\tilde{k}_n=k_n$ or $k_n^*$.}
\label{fig:syntheticdata}
\end{figure}

In Figure \ref{fig:syntheticdata} we consider two cases depending
if the shells are linearly spaced (panel a) or logarithmically
spaced (panel b). For linearly spaced shells, the curves depend on
the definitions taken for the wave number and the spectral
density. We immediately see that taking $E_n$ for the spectral
energy density leads to noisy results. The best result is obtained
taking $k_n^*$ and $E_n^{*}$.
Now taking $k_n^*$ and $E_n^{*}$ with logarithmically spaced
shells (Figure~\ref{fig:syntheticdata}(b)) leads to a result very
close to the theoretical $k^{-5/3}$ curve. Though the choice of
the logarithmically shell spacing $\lambda$ is arbitrary, we
suggest to take $\lambda=1.21$, because it is the minimum value of
$\lambda$ for which we found no empty shell.

\begin{figure}
\begin{center}
\includegraphics[width=0.49\textwidth]{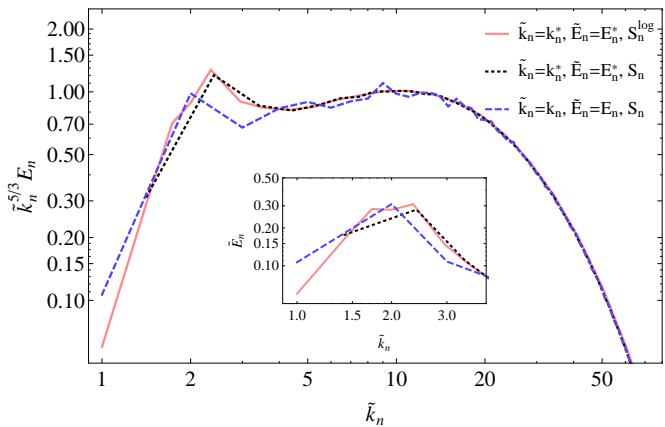}
\end{center}
\caption{Compensated spectral density $\tilde{k}_n^{5/3}\tilde{E}_n$ versus $\tilde{k}_n$ from hydrodynamic turbulent DNS data. Two curves correspond to $\tilde{k}_n=k_n$ and $\tilde{E}_n=E_n$ or $E_n^{*}$ and linearly spaced shells,
the third curve corresponds to $\tilde{k}_n=k_n^*$ and $\tilde{E}_n=E_n^{*}$ and logarithmically spaced shells.
Inlet: Spectral density at low wave numbers in the range of the forcing scales.}
\label{fig:DNSdata}
\end{figure}

Finally in Figure \ref{fig:DNSdata} we consider data from a $256^3$ direct numerical simulation of homogeneous isotropic turbulence with a random forcing \citep{Fabre2011}. The kinetic energy spectral density $E(k)$, compensated by $k^{-5/3}$, is plotted with three kinds of estimate, $k_n$ and $E_n$ (linearly spaced shells), $k_n^*$ and $E_n^{*}$ (linearly spaced shells),
$k_n^*$ and $E_n^{*}$ (logarithmically spaced shells with $\lambda=1.21$). As can be seen the curves are remarkably smooth using $k_n^*$ and $E_n^{*}$. In the inlet the spectral density $E(k)$ is plotted for low values of $k$ (without compensation). Using $E_n^{*}$, $k_n^*$ and logarithmically spaced shells (black curve), the almost constant plateau around $k=2$ corresponds indeed to the forcing scales in which energy power has been applied in the DNS.

\section{Conclusion}
\label{Conclusion}
The goal of this paper was to understand why some systematic artificial kinks appear in plots of energy spectral density issued from various DNS of 3D turbulence. We showed that they are the consequence of non self-similar distribution of Fourier modes in the spherical shells used to calculate the energy spectral density. Then we give new definitions (\ref{Eknstar})  and (\ref{knstar}) for calculating the mean energy spectral density and mean wave number in each shell. These definitions can be applied to either linearly or logarithmically spaced shells, the second one being more precise and providing a better result at low wave numbers.
The same definitions can be generalized to other scalar quantities of interest, like enstrophy, kinetic helicity in hydrodynamics, magnetic helicity and cross-helicity in magnetohydrodynamics.
Finally similar definitions can be derived for 2D problems, $S_n$ being rings instead of shells. In this case the definition of the wave number in ring $S_n$ is still given by (\ref{knstar}), but the definition of spectral energy density (\ref{Eknstar}) must be replaced by
\begin{equation}
E_n^{*}=\frac{2\pi}{M_n} \sum_{\bk'\in S_n}  \hat{E}(\bk')|\bk'|. \label{Eknstar2D}
\end{equation}

\begin{acknowledgments}

This collaboration benefited from the International Research
Group Program supported by the Perm region government. R.S. acknowledges support from
the grant YD-520.2013.2 of the Council of the President of
the Russian Federation. M.K., G.B. and F.P. acknowledge support from region Rhone-Alpes through the CIBLE program.
This work was granted access to the HPC resources of IDRIS under the allocation 20142a0611 made by GENCI and to the
supercomputer URAN of the Institute of Mathematics and Mechanics UrB RAS.

\end{acknowledgments}



  \bibliographystyle{apsrev4-1}
  \bibliography{apj-jour,refs}


\end{document}